\documentclass[prl,aps,twocolumn,footinbib,showpacs,preprintnumbers,superscriptaddress]{revtex4-1}

\usepackage{graphicx}
\usepackage{dcolumn}
\usepackage{bm}
\usepackage{amssymb}
\usepackage{pifont}
\usepackage{amsmath}
\usepackage{times}
\usepackage[nooneline]{subfigure}

\usepackage{graphicx,psfrag}
\usepackage{xcolor}

\usepackage{soul}

\begin{document}

\title{Exact mean-field theory explains the dual role of electrical synapses in collective synchronization}
\author{Ernest Montbri\'o}
\affiliation{Department of Information and Communication Technologies, 
Universitat Pompeu Fabra, 08003 Barcelona, Spain.}

\author{Diego Paz\'o}
\affiliation{Instituto de F\'isica de Cantabria (IFCA), 
CSIC-Universidad de Cantabria, 39005 Santander, Spain.}

\date{\today}

\begin{abstract}   
Electrical synapses play a major role in setting up neuronal synchronization, 
but the precise mechanisms whereby these synapses 
contribute to synchrony are subtle and remain elusive. 
To investigate these mechanisms mean-field theories 
for quadratic integrate-and-fire neurons with electrical synapses
have been recently put forward. 
Still, the validity of these theories is controversial
since they assume
that the neurons produce 
unrealistic, symmetric spikes,
ignoring the well-known impact of spike shape on synchronization.
Here we show that the assumption of symmetric spikes can be relaxed in such theories.  
The resulting mean-field equations reveal a dual role of electrical synapses: 
First, they equalize membrane potentials favoring the emergence of synchrony. 
Second, electrical synapses act as ``virtual chemical synapses", which 
can be either excitatory or inhibitory depending upon the spike shape.
Our results offer a precise mathematical explanation of the 
intricate effect of electrical synapses in collective synchronization. This 
reconciles previous theoretical and numerical works, and confirms the suitability 
of recent low-dimensional mean-field theories to investigate electrically coupled neuronal networks.
\end{abstract}
\maketitle

Electrical coupling via gap junctions is broadly present 
across brain areas~\cite{GH01,BZ04,CL04,NPR18,TWG+18}.
There is ample experimental evidence that gap junctions  
are involved in synchronizing inhibitory networks
~\cite{DTS+98,MY98,PC00,TBL00,DGS+01,HPL+01,STH+18,BB01,VLG+10,Con17}. 
Yet, and despite the fact that electrical synapses are 
recognized to constitute a critical component of the brain, 
the function of these junctions and the mechanisms whereby they 
contribute to synchrony are subtle and not 
well understood~\cite{Con17,NPR18}.  

Unlike chemical synapses, the transmission of action potentials via gap junctions
greatly depends upon the overall shape of the action potential.
This makes the effects of electrical coupling 
far from trivial, and the theoretical and computational 
analysis of electrically coupled networks of spiking neurons a difficult task.
Previous theoretical studies identified both the shape of the spikes and the 
firing frequency of the neurons as key parameters 
influencing synchrony~\cite{CK00,LR03,PMG+05,PMG+03,LS12,Boe17,OBH09}.
Yet, a precise mechanistic explanation of the role of these parameters 
in synchrony is lagging, and different works often provide 
results that are difficult to reconcile.   
For example, studies on homogeneous, two-neuron networks  
invoke the weak coupling limit and 
find that electrical and chemical synapses combine their effect 
in a linear manner~\cite{LR03,PMG+05}.
Moreover, the shape of the spikes determines whether electrical 
synapses cooperate of compete with inhibition for synchrony~\cite{PMG+05}.
In contrast, works investigating large   
heterogeneous networks suggest that electrical and inhibitory synapses 
play distinct roles: While strong electrical coupling
leads to collective synchrony~\cite{KE04,PGM+07,OBH09},
strong inhibition typically leads to the suppression of firing
and destroys synchrony~\cite{WB96,WCR+98,TJ00,DRM17}.

Rather than investigating the dynamics of large populations of spiking neurons, 
an alternative and widespread theoretical approach is to
use so-called mean-field models 
(also called firing-rate or neural-mass models)~\cite{WC72}. 
Such models are simplified, low-dimensional mathematical descriptions
of the mean activity of the population but they only characterize
populations with chemical synapses. 
Yet, an important theoretical achievement linking individual and global dynamics
has been recently accomplished with the advent of the Ott-Antonsen theory~\cite{OA08},
and with its application to populations of $\theta$-neurons~\cite{LBS13}
and Quadratic Integrate-and-Fire (QIF) neurons~\cite{MPR15}.
Notably, these novel theoretical approaches 
readily apply 
to networks  with \emph{electrical} synapses~\cite{Lai15,PDR+19,DP20,BRN+20}.

\begin{figure}[b]
\includegraphics[width=85mm,clip=true]{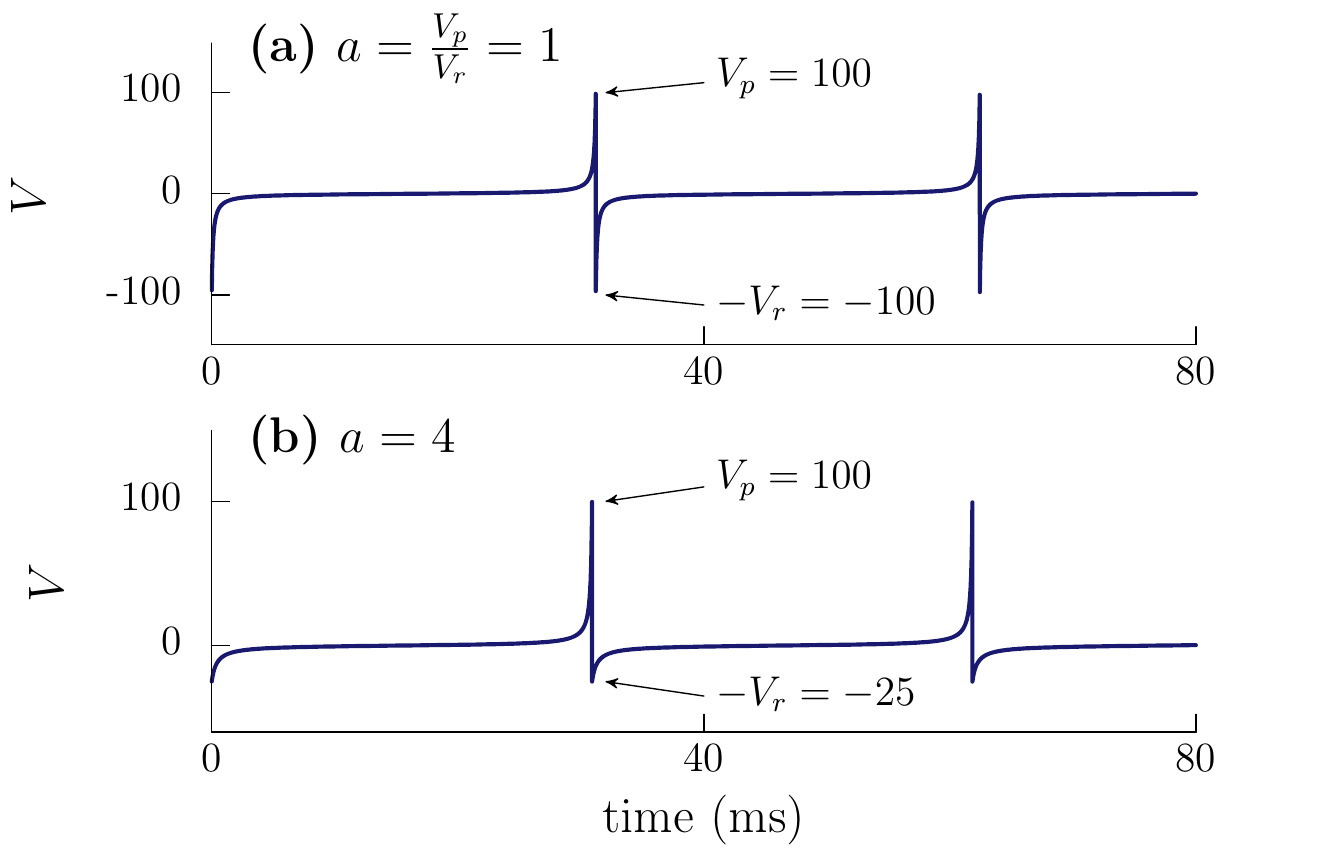}
\caption{Time series of an oscillatory 
QIF neuron, Eq.~\eqref{qif0}, with (a) symmetric, and (b) non-symmetric 
resetting rule. $\eta=1$, $\tau=10$~ms.
}
\label{Fig1}
\end{figure}

Unfortunately, the mean field theories proposed in~\cite{LBS13,MPR15} 
rely on an assumption of critical importance for electrically coupled networks: 
Spikes need to have a very particular symmetric shape 
that is unrealistic (see Fig.~\ref{Fig1}), and this may largely
alter the network dynamics. 
As we show below this is the reason why 
the low-dimensional firing-rate equations (FREs) for electrically coupled networks
originally derived in~\cite{Lai15,PDR+19} 
fail 
to elucidate how
chemical and electrical couplings  might
add their effects linearly, as found in \cite{LR03,PMG+05}.  
Therefore, though the mean field theories in~\cite{LBS13,MPR15}
have been successfully applied to investigate networks with chemical synapses
~\cite{SLB14,Lai14,RM16,PM16,RP16,DRM17,ERA+17,DEG17,DMP18,SAM+18,DT18,RP18,CB19,BAC19,DG19,KBF+19,SA20,GSK20,SBO+20,COT+20,BSV+20,BGL+20}, 
their suitability to investigate synchrony in networks with gap junctions remains questionable.

In this Letter we show that the mean-field theory originally proposed 
in~\cite{MPR15} can be generalized to account for non-symmetric spikes.
Remarkably, this leads to an extension of the FREs
in~\cite{PDR+19}, 
which provides 
the first precise mechanistic explanation 
of the dual role of electrical synapses in collective synchrony.
This explanation reconciles previous disparate theoretical and numerical 
results on the intricate effects of electrical coupling in synchrony,
and consolidates the FREs for QIF neurons as 
a valid and singular low-dimensional 
description for populations of electrically coupled neurons.

\paragraph{Quadratic integrate-and-fire neuron.--} 
First of all we introduce the QIF model, and  
explain the main idea to generalize the mean-field theory developed in~\cite{MPR15}.
The model describes the evolution of the membrane potential 
variable $V$ via the first-order ordinary
differential equation~\cite{Izh07,LRN+00,HM03}  
\begin{equation}
\tau~\dot V=V^2 + \eta;~ ~\text{if}~ V>V_p ~~\text{then}~ V \leftarrow -V_r. 
\label{qif0}
\end{equation}
The overdot denotes derivation with respect to time, 
$\tau$ is the membrane's time constant, and $\eta$ 
represents an external current. 
Due to the quadratic nonlinearity in Eq.~\eqref{qif0}, 
$V$ may escape to infinity in a finite time.
To prevent this, the QIF model 
incorporates a resetting rule:  
each time the neuron's membrane potential $V$ reaches the peak value $V_p$, 
the neuron emits a spike and the voltage is reset to  $-V_r$ (here we consider $V_p,V_r>0$), see Fig.~\ref{Fig1}. 

It is convenient to define a positive, real parameter 
characterizing spike asymmetry as the ratio
\begin{equation}
a \equiv \frac{V_p}{V_r} .
\label{a}
\end{equation}
When $a=1$,  the resetting rule of the QIF model is symmetric, see Fig.~\ref{Fig1}(a).
This symmetry is implicitly assumed in the transformation between the QIF and the $\theta$-neuron
models, which can be formally performed in the limit 
$V_r=V_p \to \infty$~\cite{EK86}.

The mean field theory in~\cite{MPR15} assumes symmetric resetting, $a=1$, and then adopts 
the limit $V_r \to \infty$. We next show that the 
assumption of symmetric resetting can be relaxed in this theory, 
leading to a novel low-dimensional firing-rate model with arbitrary spike 
asymmetry.

\paragraph{Population model.--}
We investigate a population of QIF neurons
\begin{equation}
 \tau {\dot V}_j= V_j^2+\eta_j + J \tau r(t) + g\left[v(t)-V_j\right],
\label{qif}
\end{equation}
with the resetting rule of Eq.~\eqref{qif0}. 
The index $j=1,\dots, N$ labels all the 
neurons in the population,
and the parameters $\eta_j$ represent external currents 
that are taken from some 
prescribed probability distribution $G(\eta)$.

In Eq.~\eqref{qif} electrical coupling, of strength $g\geq 0$, diffusively 
couples each neuron's membrane potential with the  mean 
membrane potential $v(t)=\frac1 N \sum_{j=1}^N  V_j(t)$.
Additionally, each neuron in the population is connected to all the other neurons  
via chemical synapses of strength $J$. 
Electrical synapses typically connect inhibitory neurons and hence $J$ should be 
thought of as a negative parameter. Chemical coupling is 
mediated by the average firing rate 
$r(t)= \frac{1}{N} \sum_{j=1}^N  \sum_{k \backslash t_j^k<t}
\int_{-\infty}^{t}dt' a_{\tau_s}(t-t')\delta (t'-t_j^{k})$, where 
$t_j^k$ is the time of the $k$th spike of $j$th neuron, $\delta (t)$ 
is the Dirac delta function, and $a_{\tau_s}(t)$
is the normalized synaptic activation caused by a single pre-synaptic 
spike (its precise shape, e.g. $a_{\tau_s}(t)=\Theta(\tau_s-t)/\tau_s$, is irrelevant as long as the time constant $\tau_s \ll 1$).

\paragraph{Theoretical analysis.--}
Next, we perform the thermodynamic limit, 
drop the indices of Eqs.~\eqref{qif},
and define the time-dependent conditional density $\rho(V|\eta,t)$, such that 
$\rho(V|\eta,t) dV$ is the fraction of neurons with membrane voltage between $V$ and $V+dV$, and
heterogeneity parameter $\eta$. 
The density $\rho$ necessarily obeys the continuity equation
\begin{equation}
\tau\partial_t \rho + \partial_V\left\{  \left[V^2+\eta+ J \tau r + g(v-V)\right] \rho \right\}=0.
\label{cont}
\end{equation}
This equation can be solved resorting to the Lorentzian ansatz
(related to the Ott-Antonsen ansatz \cite{OA08} through a conformal mapping \cite{MPR15})
\begin{equation}
\rho(V \vert \eta, t)= \frac{1}{\pi}\frac{x(\eta,t)}{\left[V-y(\eta,t)\right]^2
+x(\eta,t)^2},
\label{la}
\end{equation}
which is the asymptotic shape of the density in the long time limit
assuming that $V$ spans over the whole real line~
\cite{MPR15,PDR+19}.
We note that in numerical simulations 
$V\in(-V_r,V_p)$, and therefore
the Lorentzian ansatz Eq.~\eqref{la} is an approximation that works 
progressively better as $V_p,V_r \to \infty$.

Inserting Eq.~\eqref{la} into Eq.~\eqref{cont}, we get the evolution 
equations for $x(\eta,t)$ and $y(\eta,t)$. Nevertheless, before doing so,
it is convenient to express the global quantities $r(t)$ and $v(t)$ in Eq.~\eqref{cont}
in terms of $x(\eta,t)$ and $y(\eta,t)$. Regarding the mean firing rate $r$, 
it is related with the width $x(\eta,t)$ of the Lorentzian ansatz~\cite{MPR15}.
Indeed, the firing rate for neurons with a given $\eta$ value is 
the probability flux at $V_p$ (taken at infinity), which gives the identity:  
$r(\eta,t)= x(\eta,t)/(\pi\tau).$
Then, the mean firing rate $r(t)$ is given by the integral over all $\eta$
\begin{equation}
r(t)=\frac1{\pi\tau}\int_{-\infty}^{\infty} x(\eta,t) G(\eta)d\eta.
\label{ra}
\end{equation}

\paragraph{Mean membrane potential for general resetting rule.--}
Next we show that, for general resetting,
the mean membrane potential $v$ depends on the central voltages 
$y(\eta,t)$ as well as on the widths $x(\eta,t)$ in Eq.~\eqref{la}. 
This combined dependence underlies the 
dual effects of electrical synapses. 

The average voltage for neurons with a certain $\eta$ value 
can be calculated taking the following limit 
\footnote{The integration limits cannot be simply replaced by $\pm\infty$, 
due to the slow decay of the Lorentzian distribution $\rho(V|\eta,t) V\sim V^{-1}$ 
as $|V|\to\infty$.}
\begin{equation}
v(\eta,t)=\lim_{V_r\to\infty} \int_{-V_r}^{V_p=a V_r} \rho(V|\eta,t) \, V \, dV .
\label{veta}
\end{equation}
The mean field theory originally proposed in~\cite{MPR15} considers $a=1$. 
However, here we relax this assumption and split the integral in Eq.~\eqref{veta} 
into two parts:
one integral with symmetric integration limits (Cauchy principal value), and another one  
with the remaining integration interval
\begin{equation}
v(\eta,t)=\left[\mathrm{p.v.}\int_{-\infty}^{\infty}  + \lim_{V_r\to\infty} \int_{V_r}^{a V_r}  \right]\rho(V|\eta,t) \, V \, dV.
\label{veta2}
\end{equation}
The first integral simply yields 
the center of the distribution of membrane potentials,  $y(\eta,t)$, see Eq.~\eqref{la}. 
The second integral is the contribution to the mean membrane voltage  
due to asymmetric spike resetting, and it can be evaluated in closed form. 
 Then, after taking the limit, we find 
\begin{equation}
v(\eta,t)=y(\eta,t)+ \frac{\ln a}\pi  x(\eta,t) . 
\label{v0}
\end{equation}
This identity pinpoints the deviation of the 
mean membrane potential, $v(\eta,t)$, with respect to
the center of the distribution of voltages, $y(\eta,t)$,
due to asymmetric spike resetting, $a\neq 1$. Remarkably, this 
deviation is both proportional to the firing rate through $x(\eta,t)$, 
and to  $\ln a$. Therefore, for spikes with $a>1$, the mean membrane voltage is 
above the center of the distribution, whereas for spikes with $0<a<1$ 
the mean membrane voltage is below $y(\eta,t)$.

For symmetric resetting, the mean membrane potential $v_s(t)$ of the entire population is 
obtained integrating $y(\eta,t)$ over all $\eta$ values:
\begin{equation}
v_s(t) \equiv \int_{-\infty}^{\infty}  y(\eta,t) G(\eta)  d\eta .
\label{vs}
\end{equation}
In the general case of asymmetric spike resetting, the mean 
membrane voltage is obtained integrating Eq.~\eqref{v0} over $\eta$. 
Hence, using Eqs.~\eqref{ra} and \eqref{vs}, we find
\begin{equation}
v(t)=  v_s(t) + (\tau \ln a )~r(t).
\label{v}
\end{equation}

 \paragraph{Dynamics in the ``Lorentzian manifold''.--}
The evolution equations for $x(\eta,t)$ and $y(\eta,t)$, 
obtained substituting Eq.~\eqref{la} into Eq.~\eqref{cont}, 
can be condensed into one via the complex variable $w(\eta,t)\equiv x(\eta,t) +i y(\eta,t)$:
\begin{equation}
\tau\partial_t w(\eta,t)= i\left[\eta + (J+g \ln a) \tau r -w^2\right] + 
g(i v_s -w), 
\label{w}
\end{equation}
where $r$ and $v_s$ are the integrals in Eqs.~\eqref{ra} and \eqref{vs}.
Note that Eq.~\eqref{w} is an infinite dimensional system, since each $w(\eta,t)$ 
is coupled to all the other functions $w(\eta',t)$. 

Three results
follow from Eq.~\eqref{w},
which are valid irrespective of the specific 
form of the distribution $G(\eta)$:
\emph{First}: Only in the presence of electrical coupling ($g> 0$), 
the effects of asymmetric spike resetting influence the collective dynamics of the network. 
\emph{Second}: 
The term $\ln a$ weights the 
`virtual chemical coupling' that linearly adds to the
actual chemical coupling constant $J$. Moreover, the sign of $\ln a$ determines
if this contribution is excitatory or inhibitory. 
Thus, in the presence of electrical coupling, the 
\emph{effective} chemical coupling constant becomes 
\begin{equation}
J_{\text{eff}}(J,g,a)=J+ g \ln a.
\label{Jeff}
\end{equation}
\emph{Third}: Terms with electrical coupling $g$ appear at two different places 
in Eq.~\eqref{w}, indicating a dual role of gap junctions in the dynamics of $w$.

\begin{figure}
\includegraphics[width=85mm,clip=true]{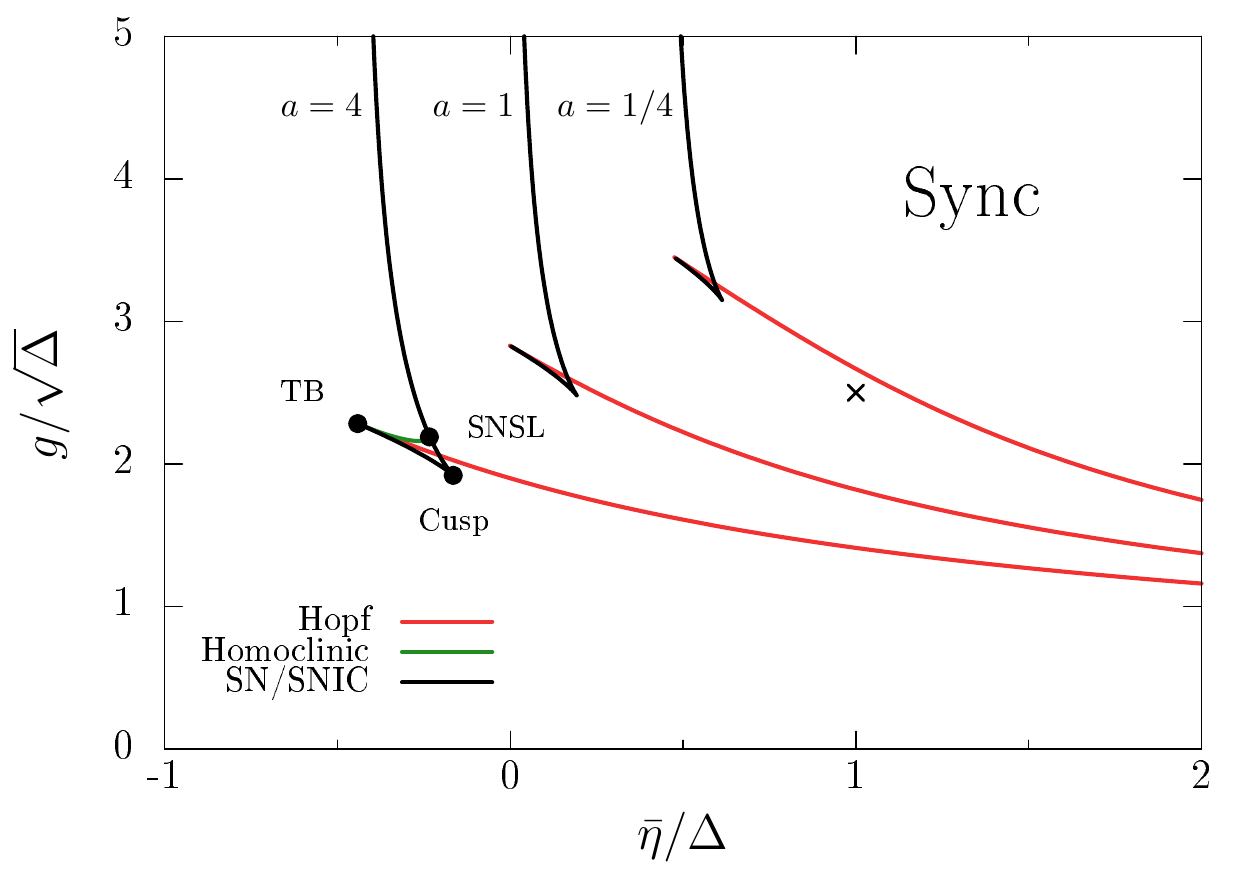}
\caption{Phase diagram of the FREs~\eqref{fre} for electrical coupling
only ($J=0$) and for three different values of the asymmetry parameter $a$.
The region of synchronization (`Sync'), located 
 in the upper-right part of the diagram, enlarges as $a$ grows.
Symbol $\times$ indicates the parameters value $(1,2.5)$,   
used in Fig.~\ref{Fig3}. To lighten de diagram,  
homoclinic bifurcations and codimension-2 points 
---Cusp, Saddle-Node-Separatrix-Loop (SNSL), and
Takens-Bogdanov (TB)---, are omitted in the cases $a=1$ and $a=1/4$.  
}
\label{Fig2}
\end{figure}

\paragraph{Firing-rate equations.--} 
To gain further insight on the effects of electrical coupling
we adopt hereafter Lorentzian distributed currents 
with center at $\bar \eta$ and half-width $\Delta$: $G(\eta)=(\Delta/\pi)/[(\eta-\bar\eta)^2+\Delta^2]$.
In this case a maximal dimensionality reduction is achieved, 
since the integrals in Eqs.~\eqref{ra} and \eqref{vs}
can be evaluated applying the residue theorem. 
We analytically extend $w(\eta,t)$ into complex $\eta$ \cite{MPR15}, close
the integrals by an arc at infinity in the complex half-plane $\text{Im}(\eta)<0$ 
\footnote{The integrals 
along the arc at infinity are zero,
since according to Eq.~\eqref{w} we have: $|w|\sim\sqrt{|\eta|}$. 
Accordingly, e.g.~for Eq.~\eqref{ra}, 
we have $\int_{0}^{-\pi} \mathrm{Re}(w)  G(\eta) |\eta| i e^{i\phi} d\phi\sim |\eta|^{3/2}/|\eta|^2\to 0$; likewise for Eq.~\eqref{vs}.},
and obtain: 
$\pi\tau r(t) + i v_s= w(\bar\eta-i\Delta,t)$.
Then, we evaluate Eq.~\eqref{w} at $\eta=\bar\eta-i\Delta$,
obtaining the FREs
\begin{subequations}
\label{fre}
\begin{eqnarray}
\tau \dot r &=& \frac{\Delta}{\tau \pi} + 2  r v_s - g r,  \label{rdot}\\
\tau \dot v_s &=&   v_s^2 +   \bar \eta - (\pi \tau r)^2 + (J+g \ln a ) \tau r.
\label{vdot}
\end{eqnarray}
\end{subequations}
%
\begin{figure*}
\includegraphics[width=180mm,clip=true]{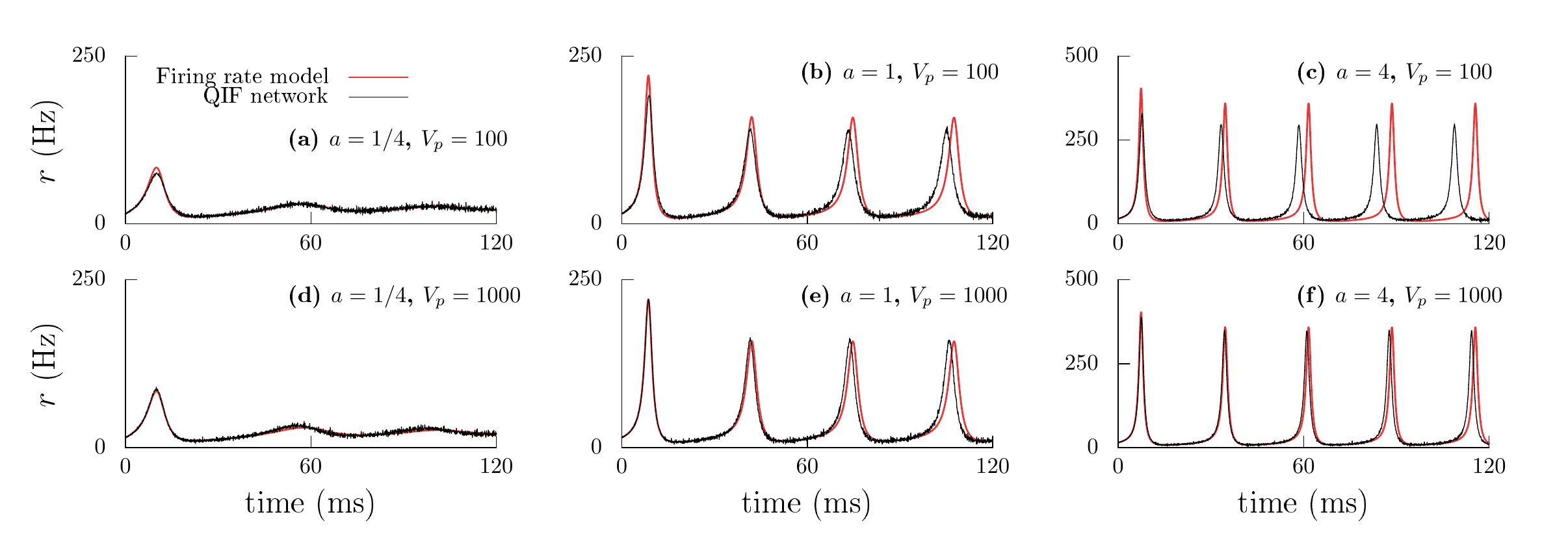}
\caption{Time series of $r(t)$ for FREs~\eqref{fre} (red) and 
network Eqs.~\eqref{qif} with $N=10^4$ (black). Peak values are (a,b,c) $V_p=100$,  and (d,e,f)  $V_p=1000$. 
Three values of the spike asymmetry parameter are used: 
(a, d) $a=1/4$, (b,e) $a=1$, (c,f) $a=4$. 
Parameters correspond to the symbol $\times$ in Fig.~\ref{Fig2}: $J=0$, $g=2.5$, $\bar\eta=\Delta=1$, and $\tau=10$~ms. We used the explicit Euler scheme with  $dt=10^{-4}$
ms and $\tau_s=10^{-3}$ ms.
}
\label{Fig3}
\end{figure*}
This system of two ordinary differential equations
for the mean firing rate $r$ and for the auxiliary variable $v_s$
describes the dynamics of the ensemble exactly in the limit of infinite  $N$, $V_p$ and $V_r$.
\footnote{It is also possible to write Eqs.~\eqref{fre} 
using the variables $r$ and $v$. Here we use the auxiliary variable $v_s$ instead of 
$v$ since in this form the FREs are simpler to interpret.}. 
The dual role of the electrical coupling $g$ is clearly described by the FREs: 
First, electrical coupling enters in Eq.~\eqref{rdot},
reducing the firing rate, and hence reducing the width of the 
distribution of membrane potentials. This contribution homogenizes membrane potentials 
and, for large enough $g$, favors the emergence of collective synchronization 
\footnote{For symmetric resetting, $a=1$, 
the mean membrane voltage is $v=v_s$ and the FREs derived in~\cite{PDR+19} are recovered. 
In this case chemical and 
electrical coupling act independently, in different terms of Eqs.~\eqref{fre}.}.
Second, electrical coupling enters in Eq.~\eqref{vdot}, contributing as 
a virtual chemical coupling. Depending on the 
value of $a$, electrical and 
chemical synapses cooperate or compete for synchrony.

The relevant parameters of Eqs.~\eqref{fre} are $\bar\eta$, 
the coupling constants $J$ and $g$, and the spike asymmetry $a$ 
(one can assume  
$\Delta=1$ and $\tau=10$~ms
without loss of generality). 
Additionally, parameters $a$ and $J$ are both acting in the same term of 
Eq.~\eqref{fre} and hence they can not produce qualitatively different dynamical behaviors. 
Accordingly, we consider $J=0$ 
hereafter
and borrow the results of the 
bifurcation analysis in~\cite{PDR+19}, which can be directly applied here 
replacing the chemical coupling 
constant $J$ by the 
effective chemical coupling Eq.~\eqref{Jeff} 
\footnote{The weakly nonlinear analysis performed in~\cite{PDR+19} 
also applies here, and confirms that ---in contrast to \cite{OBH09}---
the Hopf bifurcation remains always 
supercritical for any value of $a$.}.
The phase diagram in Fig.~\ref{Fig2} shows  the 
bifurcation loci of the FREs for three values of $a$. 
Three bifurcations, Hopf (red), Saddle-Node (black) and Homoclinic (green), 
tangentially meet at the Takens-Bogdanov (TB) point, 
\begin{equation}
\left(\bar \eta/\Delta \right)_{\text{TB}}= -\tfrac{1}{\pi}\ln a  ,
\label{TB}
\end{equation}
around which the phase diagram organizes. The homoclinic curve 
(shown only for $a=4$) 
moves parallel to the Hopf line for a while, and then   
tangentially meets the upper branch of the Saddle-Node (SN) bifurcating line 
at a saddle-node-separatrix-loop (SNSL) point. 
At this point, the SN boundary becomes a SN on the Invariant Circle (SNIC) boundary (black) 
that together with the Hopf and 
homoclinic lines encloses the region of synchronization (Sync) where 
collective oscillations occur. Note that two small regions of bistability are located 
in the region limited by the three codimension-two points: Cusp, TB, and SNSL.

In Fig.~\ref{Fig2} the SNIC bifurcation lines
have a vertical asymptote 
precisely at the same $\bar\eta$ value than
the TB point Eq.~\eqref{TB}. 
For $0<a<1$, the asymptote is located at positive values of 
$\bar\eta$, and  shifts to the left as $a$ increases. 
Hence, synchronization is 
hindered for spikes with $0<a<1$, 
and favored for spikes with $a>1$~
\footnote{The asymptote of the SNIC boundary is located at $\bar \eta<0$
for $a>1$. Therefore, even if a majority of the neurons are intrinsically quiescent, 
electrical coupling is able to trigger
a self-sustained collective
oscillation. Similar effects have been reported for pairs of 
electrically coupled neurons~\cite{MRS+97,KMA90}.}.
In Fig.~\ref{Fig3} we depict the time series of the firing rate in numerical simulations 
of both the network Eqs.~\eqref{qif} and the FREs Eqs.~\eqref{fre},
for three values of the parameter $a=\{1/4,1,4\}$. In the first row, we used spikes with 
$V_p=100$ in  Eqs.~\eqref{qif}. Though the agreement between the models 
is not perfect, the mean field Eqs.~\eqref{fre} accurately 
predict the emergence of oscillations in Eqs.~\eqref{qif} as $a$ is increased. 
Finally, for $V_p=1000$ (second row), 
the agreement greatly improves showing that the dynamics of the network Eqs.~\eqref{qif}  
converges to that of the FREs for large values of $V_p$.

\paragraph{Conclusions.--} We showed that 
the spike resetting rule of 
the QIF model can be incorporated in the 
mean-field theory proposed in~\cite{MPR15}.
This extension of the theory 
reveals 
a nontrivial dependence of the mean membrane voltage
on the firing rate ---see Eq.~\eqref{v}--- which turns out to be crucial for 
deciphering the dual role of electrical synapses in synchrony, 
and to reconcile previous theoretical and numerical work. 
As the dynamics in the ``Lorentzian manifold'' Eq.~\eqref{w}
as well as the FREs~\eqref{fre} are valid for 
arbitrary
coupling 
strengths, we conclude
that electrical and chemical coupling do not 
simply
add linearly 
for weak coupling, as suggested by previous works, see e.g.~\cite{KE04}.
In view that traditional mean-field theories for neural networks only encompass 
 chemically coupled ensembles, 
we regard the mean-field theory presented here
as a unique tool to investigate the interplay 
between chemical and electrical synapses.

\acknowledgments
We acknowledge support by the Agencia Estatal de Investigaci\'on and 
Fondo Europeo de Desarrollo Regional under Project
 No.~FIS2016-74957-P (AEI/FEDER, EU).

%


 \makeatletter
\setcounter{equation}{0}
\renewcommand{\theequation}{A\arabic{equation}}
 \makeatother

\end{document}